\documentclass[twocolumn,twoside,superscriptaddress,prl,aps]{revtex4}
\usepackage{amsmath,mathrsfs,amsbsy,color,graphicx,bm,amsthm,amsfonts}
\usepackage{units}
\usepackage{bbm}
\usepackage{color,ulem}
\usepackage[colorlinks,citecolor=blue]{hyperref}

\newcommand{\tr}{{\rm Tr}}

\newcommand{\ket}[1]{|#1\rangle}

\newcommand{\B}[1]{\textcolor{blue}{#1}}

\begin{document}
\title{Realization of the tradeoff between internal and external entanglement}
\author{Jie Zhu}
\affiliation{Key Laboratory of Quantum Information, University of Science and Technology of China, CAS, Hefei, 230026, China}
\affiliation{CAS Center for Excellence in Quantum Information and Quantum Physics, Hefei, 230026, China}
\author{Meng-Jun Hu}
\affiliation{Key Laboratory of Quantum Information, University of Science and Technology of China, CAS, Hefei, 230026, China}
\affiliation{CAS Center for Excellence in Quantum Information and Quantum Physics, Hefei, 230026, China}
\author{Yue Dai}
\affiliation{School of Physical Science and Technology, Soochow University, Suzhou, 215006, China}
\author{Yan-Kui Bai}
\affiliation{College of Physics Science and Information Engineering and Hebei Advanced Thin Films Laboratory,
Hebei Normal University, Shijiazhuang, Hebei 050024, China}
\author{S. Camalet}
\affiliation{Sorbonne Universit\'{e}, CNRS, Laboratoire de Physique Th\'{e}orique de la Mati\`{e}re Condens\'{e}e,
LPTMC, F-75005 Paris, France}
\author{Chengjie Zhang}
\email{chengjie.zhang@gmail.com}
\affiliation{School of Physical Science and Technology, Soochow University, Suzhou, 215006, China}
\author{Chuan-Feng Li}
\affiliation{Key Laboratory of Quantum Information, University of Science and Technology of China, CAS, Hefei, 230026, China}
\affiliation{CAS Center for Excellence in Quantum Information and Quantum Physics, Hefei, 230026, China}
\author{Guang-Can Guo}
\affiliation{Key Laboratory of Quantum Information, University of Science and Technology of China, CAS, Hefei, 230026, China}
\affiliation{CAS Center for Excellence in Quantum Information and Quantum Physics, Hefei, 230026, China}
\author{Yong-Sheng Zhang}
\email{yshzhang@ustc.edu.cn}
\affiliation{Key Laboratory of Quantum Information, University of Science and Technology of China, CAS, Hefei, 230026, China}
\affiliation{CAS Center for Excellence in Quantum Information and Quantum Physics, Hefei, 230026, China}

\begin{abstract}
We experimentally realize the internal and external entanglement tradeoff, which is a new kind of entanglement monogamy relation different from that usually discussed. Using a source of twin photons, we find that the external entanglement in polarization of twin photons, and the path-polarization internal entanglement of one photon, limit each other. In the extreme case, when the internal state is maximally entangled, the external entanglement must be vanishing, that illustrate entanglement monogamy. Our results of the experiment coincide with the theoretical predictions, and therefore provide a direct experimental observation of the internal and external entanglement monogamy relation.
\end{abstract}
\date{\today}

\maketitle

Entanglement monogamy is one of the most fundamental properties for multipartite quantum states, which means that if two qubits $A$ and $B$ are maximally entangled, then $A$ or $B$ cannot be entangled with the third qubit $C$ \cite{mono1,mono2}. The quantitative entanglement monogamy inequality was first proved by Coffman, Kundu, and Wootters (CKW) for three-qubit states \cite{CKW},
\begin{equation}\label{CKW}
    C^2_{A|B}+C^2_{A|C}\leq C^2_{A|BC},
\end{equation}
where $C^2$ denotes the squared concurrence for quantifying bipartite entanglement \cite{Wootters}. From Eq. (\ref{CKW}), one can easily find that there is a consequent tradeoff between the amount of entanglement shared by qubits $A$ and $B$, and the entanglement shared by qubits $A$ and $C$. For three-qubit pure states, the difference between the right hand side and left hand side of Eq. (\ref{CKW}) is defined as the so-called ``three-tangle" \cite{CKW}, which is a genuine three-qubit entanglement measure. After the CKW inequality, several entanglement monogamy inequalities \cite{NCKW,gaussian1,gaussian2,gaussian3,gaussian4,SEntangle1,SEntangle2,negativity,CREN,Kim2,Allen,SEOF,SEOF2,SEOF3,bai2,Kim3,Renyi,Kim4,song,song2,exp,exp2,
Gour,Yu,Sen,Sen2,Fei,Fei2,Fei3,ou2,Regula,Regula2,Osterloh,Luo,Luo2,Winter,Eltschka,Eltschka2,Eltschka3} and even monogamy equalities \cite{equa,equa2} were introduced. Osborne and Verstraete proved the CKW monogamy inequality for $N$-qubit states \cite{NCKW}. In Refs. \cite{gaussian1,gaussian2}, the CKW inequality was generalized to Gaussian states. Moreover, other entanglement measures, such as the squashed entanglement \cite{SEntangle1,SEntangle2}, the negativity \cite{negativity,CREN,Kim2,Allen}, and the squared entanglement of formation \cite{ SEOF,SEOF2,SEOF3}, were also employed to derive the corresponding entanglement monogamy inequalities.

\begin{figure}
\includegraphics[scale=0.27]{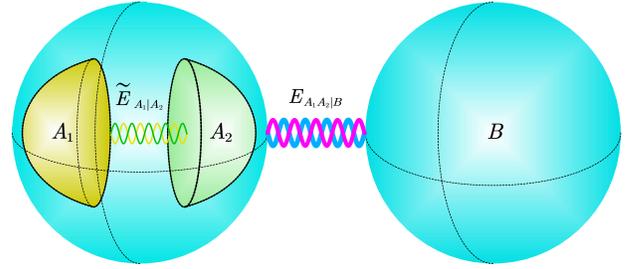}
\caption{For a tripartite quantum state $\varrho_{A_1 A_2 B}$, the subsystems $A_1$ and $A_2$ are in the same physical system but they are encoded in different degrees of freedom, and $B$ is encoded in another physical system. $\tilde{E}_{A_1|A_2}$ and $E_{A_1 A_2|B}$ represent the internal entanglement between $A_1$ and $A_2$ and external entanglement between $A_1A_2$ and $B$, respectively.}\label{fig1}
\end{figure}

Recently, new kinds of monogamy relation have been derived by Camalet \cite{Camalet,Camalet2,Camalet3,Camalet4}, i.e. internal entanglment (or local quantum resource) and external entanglement have a tradeoff. The usually discussed entanglement monogamy inequalities in Refs. \cite{NCKW,gaussian1,gaussian2,SEntangle2,negativity,CREN,Kim2,SEOF,SEOF2,SEOF3} indicate the trade-off relation between $E(\varrho_{AB})$ and $E(\varrho_{AC})$ (or its extension to $N$-partite case), where $E$ is one kind of entanglement measure, $\varrho_{AB}$ and $\varrho_{AC}$ are reduced density matrices from a three-qubit state. Unlike these previously derived inequalities, Camalet has proposed a new entanglement monogamy inequality \cite{Camalet}. Consider a tripartite quantum state $\varrho_{A_1 A_2 B}$ illustrated in Fig. \ref{fig1}, where $A_1$ and $A_2$ come from the same physical system but have been encoded in different degrees of freedom, and $B$ is encoded in another physical system. This inequality shows the tradeoff relation between the internal entanglement $\tilde{E}_{A_1|A_2}$ and the external entanglement $E_{A_1 A_2|B}$, where $\tilde{E}$ and $E$ are two different but related entanglement measures, and $E_{A_1 A_2|B}$ ($\tilde{E}_{A_1|A_2}$) denotes the entanglement of $\varrho_{A_1 A_2 B}$ ($\varrho_{A_1 A_2}$) under the bipartition $A_1 A_2|B$ ($A_1|A_2$).

Here we experimentally demonstrate the entanglement monogamy relation between the internal and external entanglement, with a source of twin photons. As shown in Fig. \ref{fig1}, there are two qubits (the polarization qubit $A_1$ and the path qubit $A_2$) encoded in photon $A$; but only one qubit (the polarization qubit $B$) is encoded in photon $B$. Here we provide a direct experimental observation of the tradeoff between the internal entanglement in $A_1|A_2$ and the external entanglement in $A_1A_2|B$.

\textit{Theoretical framework.---} Let us focus on a tripartite state $\varrho_{A_1 A_2 B}$ 
where $A_1$ and $A_2$ are encoded in the same physical system $A$ by using different degrees of freedom, see Fig. \ref{fig1}. The third party is encoded in system $B$. Camalet's entanglement monogamy inequality is 
\begin{equation}\label{monogamy}
    \tilde{E}_{A_1|A_2}+E_{A_1A_2|B}\leq \tilde{E}_{\max},
\end{equation}
where $\tilde{E}_{A_1|A_2}$ denotes the internal entanglement measure between $A_1$ and $A_2$, $E_{A_1A_2|B}$ is the external entanglement measure between $A_1A_2$ and $B$, and $\tilde{E}_{\max}$ is the 
value of  $\tilde{E}_{A_1|A_2}$ when $A_1$ and $A_2$ are maximally entangled \cite{Camalet}. It is worth noting  that $\tilde{E}$ and $E$ are strongly related, although they are two different entanglement measures. From  the inequality (\ref{monogamy}), one can see that $E_{A_1A_2|B}$ is also bounded by $\tilde{E}_{\max}$. 
When the state $\varrho_{A_1 A_2 B}$ is pure and the reduced density operator $\varrho_{A_1 A_2}$ is absolutely separable \cite{AS,MEMS}, the external entanglement $E_{A_1A_2|B}$ is equal to the maximum value $\tilde{E}_{\max}$.
On the other hand, when the internal entanglement $\tilde{E}_{A_1|A_2}$ is maximal, the external entanglement $E_{A_1A_2|B}$ must be vanishing.

\begin{figure}
\includegraphics[width=7.5cm]{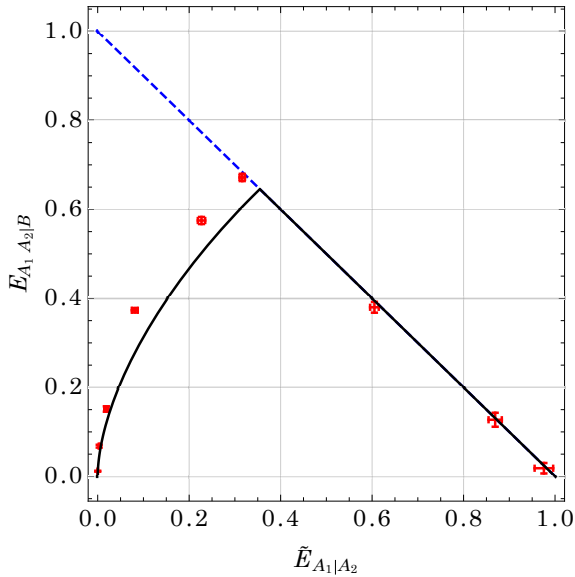}
\caption{The internal entanglement $\tilde{E}_{A_{1}|A_{2}}$ quantified by $E_F(\varrho_{A_1 A_2})$ and the external entanglement $E_{A_{1}A_{2}|B}$ quantified by $E_F'(|\psi\rangle_{A_1 A_2 |B})$ are bounded by the dashed blue straight line. We calculated the states in Eq. (\ref{thstate}) and the corresponding nine different experimental states, the results are shown as the black curve and red dots, respectively. The left six red dots should be on the curve according to the theoretical prediction but some dots do not. The deviation is from the visibility of interferometers, and the error bar is from the Poissonian distribution of photon counts.}\label{fig2}
\end{figure}

The internal entanglement measure $\tilde{E}_{A_1|A_2}$ in Eq. (\ref{monogamy}) can be arbitrary entanglement measures, such as the entanglement of formation $E_F$ \cite{Wootters}, the negativity $E_N$ \cite{neg1,neg2}, and the relative entropy of entanglement $E_R$ \cite{ER}. When we choose the entanglement of formation $E_F$ to quantify the internal entanglement between $A_1$ and $A_2$, the inequality (\ref{monogamy}) for a general three-qubit pure state $|\psi\rangle_{A_1 A_2 B}$ becomes to
\begin{eqnarray}
E_F(\varrho_{A_1 A_2})+E_F'(|\psi\rangle_{A_1 A_2 |B})\leq1,\label{4}
\end{eqnarray}
where the internal entanglement $\tilde{E}_{A_{1}|A_{2}}$ is $E_F(\varrho_{A_1 A_2})=H\big(1/2+\sqrt{1-C^2(\varrho_{A_1 A_2})}/2\big)$, $H$ is the binary entropy $H(x):=-x \log_2 x-(1-x)\log_2 (1-x)$, and $C(\varrho)=\max\{0,\sigma_1-\sigma_2-\sigma_3-\sigma_4\}$ is the concurrence of $\varrho$ with $\{\sigma_i\}$ being the square roots of eigenvalues of $\varrho\sigma_y\otimes\sigma_y \varrho^*\sigma_y\otimes\sigma_y$ in decreasing order  \cite{Wootters}. The external entanglement $E_{A_{1}A_{2}|B}$ is $E_F'(|\psi\rangle_{A_1 A_2 |B})$, as defined by

\begin{eqnarray}
E_F'(|\psi\rangle_{A_1 A_2 |B})&:=&1-\max_U E_F(U\varrho_{A_1 A_2}U^\dag)\label{5}\\
&=& 1- f(\max\{0,\lambda_1-\lambda_3-2\sqrt{\lambda_2\lambda_4}\})\nonumber
\end{eqnarray}
where $U$ denotes the unitary operators of $A$,
$f(x):=H(1/2+\sqrt{1-x^2}/2)$, and $\{\lambda_i\}$ are the eigenvalues of $\varrho_{A_1 A_2}$ in nonascending order \cite{Camalet,MEMS}. In Ref. \cite{MEMS}, the maximum entanglement 
for a given spectrum $\{\lambda_i\}$ measured by the negativity and the relative entropy of entanglement have also been provided. Thus, one can obtain the inequality (\ref{monogamy}) with the internal entanglement measure being the negativity and the relative entropy of entanglement as well \cite{SM}.

\begin{figure*}
\includegraphics[width=16cm]{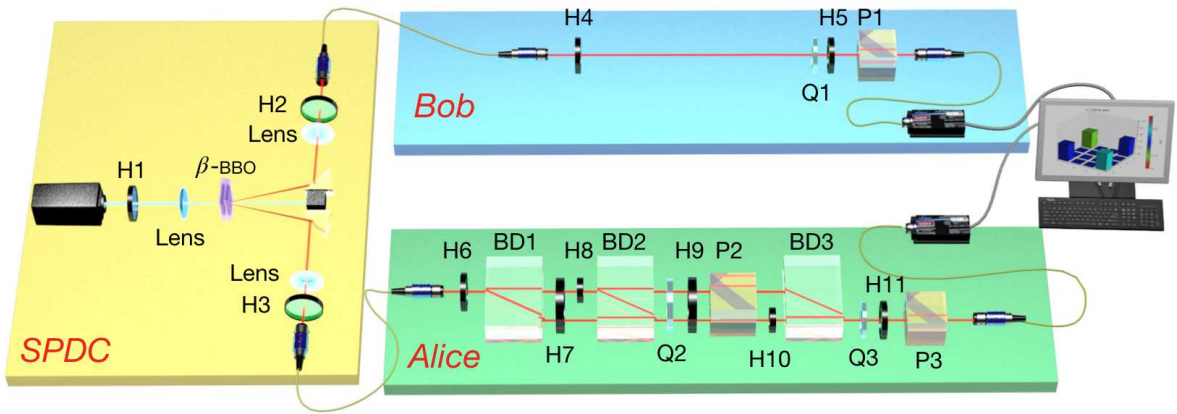}
\caption{\textbf{Experimental setup} The polarization entangled photon pairs are generated by the spontaneous parameteric down-conversion process. In the Alice part, the polarization and path states are entangled. In each mode, half-wave plate (HWP), quarter-wave plate (QWP) and polarization beam splitter (PBS) are set for state tomography. In the experiment, the photons are collected by two single photon counting modules and identified by the coincidence counter. H: half wave plate H1$\sim$H11; Q: quarter wave plate Q1$\sim$Q3; P: polarization beam splitter P1$\sim$P3; BD: beam displacer BD1$\sim$BD3.}\label{fig3}
\end{figure*}

Now we consider a class of three-qubit pure states with one parameter $\phi$,
\begin{eqnarray}
|\psi\rangle=\cos\phi |110\rangle+\sin\phi \frac{|01\rangle+|10\rangle}{\sqrt{2}}|1\rangle.\label{thstate}
\end{eqnarray}
Based on Eqs. (\ref{4})-(\ref{5}), one can obtain its internal and external entanglement by using the entanglement of formation,
\begin{eqnarray}
E_F(\varrho_{A_1 A_2})&=&   f(\sin^2\phi),\\
E_F'(|\psi\rangle_{A_1 A_2 |B})&=& 1-f(\max\{\cos^2\phi,\sin^2\phi\}).
\end{eqnarray}
The theoretical results have been shown in Fig. \ref{fig2} by the solid curve. We can see that all the results are bounded by the dashed line, i.e., the monogamy inequality (\ref{4}) always holds.

\textit{Experimental realization.---}
In order to demonstrate this new entanglement monogamy relation, we prepare some quantum states where the quantity of internal and external entanglement can be controlled. We use the polarization and the path degrees of freedom to produce the target three-qubit state in Eq. (\ref{thstate}).

As shown in Fig. \ref{fig3}, we will introduce three parts of the setup: (i) state preparation, (ii) qubits $A_1$ and $A_2$ (owned by Alice), (iii) qubit $B$ (owned by Bob). First, the source of twin photons is realized via a type-I spontaneous parametric down-conversion (SPDC) process in which the crystal is a joint $\beta$-barium-borate ($\beta$-BBO) \cite{P.G.Kwiat1999}. The source of the two-qubit entangled state is $|\psi\rangle=\cos\phi\left|H\right\rangle_{A}\left|H\right\rangle_{B}+\sin\phi\left|V\right\rangle_{A}\left|V\right\rangle_{B}$, where the parameter $\phi$ is modulated by H1, a half-wave plate (HWP) put in front of the BBO crystal to adjust the polarization of pump. Here the pump is a continuous-wave diode laser with 140mW and the wavelength is 404nm. The fidelity between the experimental state and the theoretical state is beyond 99$\%$. The computational basis {$\left|0\right\rangle$ and $\left|1\right\rangle$} are encoded in the horizontal polarization $\left|H\right\rangle $ and vertical polarization $\left|V\right\rangle$ of the photons, respectively. The photon pair is separated spatially via a single mode fiber (SMF). One photon is sent to Alice and the other one is sent to Bob. In Alice part, as illustrated in Fig. \ref{fig3}, we use three beam displacers (BDs) in which the vertical-polarized photon remains on its path while the horizontal-polarized photon shifts down. The internal entanglement is realized between the path and the polarization degrees of freedom of Alice's photon. The upper path state is encoded into $\left|0\right\rangle$, and the down path state is corresponding to $\left|1\right\rangle$. After BD1, the photons in different polarization states travel two paths. Thus the polarization and path are entangled.
Due to the HWP at $45^{\circ}$ (H7), the horizontal and vertical polarization exchange, whereafter the HWP in the upper path (H8) rotates the polarization. The angle $\theta$ modulated by H8 is the controllable parameter of the internal entanglement. Right after the H8, there is another beam displacer, BD2, to fulfill the preparation of the internal entanglement of the Alice's photon. On the other hand, in Bob's part, the photons are measured directly. Finally we get the three-qubit states which contain internal and external entanglement and can be described as
\begin{eqnarray}
 |\psi\rangle=\cos\phi |1\rangle_{A_{1}}| 1 \rangle _{A_{2}}| 0\rangle_{B} +\sin\phi|\varphi\rangle_{A_1 A_2}|1\rangle_{B},\label{estate}
\end{eqnarray}
where $|\varphi\rangle_{A_1 A_2}=\cos\theta| 0\rangle_{A_{1}}| 1 \rangle  _{A_{2}} +\sin\theta| 1\rangle_{A_{1}}| 0 \rangle _{A_{2}}$.

In order to reconstruct the density matrices of these three-qubit states, we perform quantum state tomography for these states. According to the maximum likelihood estimation, the density matrices are reconstructed \cite{Daniel F.V.James2001}. The project measurement of polarization is realized by a standard polarization tomography setup (SPTS), which consists of a quart-wave plate (QWP), a half-wave plate, and a polarization beam splitter (PBS). As shown in the Fig. \ref{fig3}, there are three such setups, (Q1,H5,P1), (Q2,H9,P2), and (Q3,H11,P3). The first two setups measure the polarization states of Bob and Alice respectively. As for the last one, it is used to measure the path state of Alice. The following is demonstrated how it works. Consider the BD2, it is the last element in the state preparation. If a photon is in the upper path after BD2, its polarization is horizontal when it arrives at Q3, i.e. the last SPTS; and if it is in the down path, it will be vertical-polarized. Therefore, the last SPTS measure the path state via the polarization state tomography. Besides it is necessary to mention that the HWP right after P2 in down path (H10) is at  $ 45^{\circ}$.

\begin{figure}
\includegraphics[width=8cm]{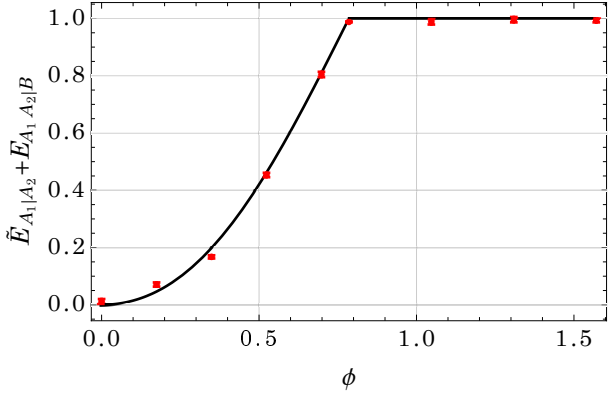}
\caption{We adjusted the degree of H1 to change the values of $\phi$ and chose nine states to calculate $\tilde{E}_{A_{1}|A_{2}}+E_{A_{1}A_{2}|B}$. The red dots are the experimental results and the black curve is the theoretical values. The error bar is from the Poissionian distribution of photon counts.}\label{fig4}
\end{figure}

We fix the value of $\theta$ at $45^\circ$ and adjust the H1 to change the values of $\phi$ from $0^\circ$ to $90^\circ$. Thus, the experimental state in Eq. (\ref{estate}) becomes to Eq. (\ref{thstate}).
If $\phi=90^\circ$, the internal entanglement of system $A$ is maximal; and if $\phi=0^\circ$, the state is $|\psi\rangle=|110\rangle$ which is separable. Without loss of generality, we choose nine typically states to analyze. We reconstruct the density matrices of these nine states, then calculate their entanglements $\tilde{E}_{A_{1}|A_{2}}$ and $E_{A_{1}A_{2}|B}$ based on $E_F$ using the expressions given above. As shown in Fig. \ref{fig4}, the experimental results coincide with the theoretical prediction within the margin of error. Although the different value of the angle $\phi$ represent the different states, the sum of $\tilde{E}_{A_{1}|A_{2}}+E_{A_{1}A_{2}|B}$ will not exceed 1, 
which experimentally demonstrates the monogamy relation 
\eqref{monogamy}. We remark that $E_{A_{1}A_{2}|B}$ is evaluated using
Eq. \eqref{5} which is strictly speaking valid only for genuine pure states. However,
the actual entanglement $E_{A_{1}A_{2}|B}$ of the experimental state, which is
not exactly pure, is lower than the value obtained from Eq. \eqref{5} \cite{Camalet,Camalet3,SM}.
On the other hand, we choose $\tilde{E}_{A_{1}|A_{2}}$ as the horizontal ordinate and $E_{A_{1}A_{2}|B}$ as the vertical ordinate to plot Fig. \ref{fig2}. We find that $E_{A_{1}A_{2}|B}$ firstly increases and then decreases as $\tilde{E}_{A_{1}|A_{2}}$ increases, which also agrees with the theoretical results. Moreover, these values are in the area below the straight line $E_{A_{1}A_{2}|B}=1-\tilde{E}_{A_{1}|A_{2}}$.

\textit{Other monogamy inequalities.---}
In Ref.~\cite{Camalet}, Camalet also presented a monogamy inequality involving only one entanglement monotone, the negativity $E_N$. For a bipartite state $\varrho_{AB}$, the negativity is defined by $E_N(\varrho_{AB})=(\|\varrho_{AB}^{T_B}\|-1)/2$ \cite{neg1,neg2}, where $\|\cdot\|$ is the trace norm and $T_B$ is the partial transpose 
with respect to system $B$. Contrary to other entanglement measures, such as
$E_F$ or $E'_F$, $E_N$ is readily computable for any state.
For a three-qubit state $\varrho_{A_1A_2B}$, the monogamy inequality is
\begin{equation}\label{EN}
E_N(\varrho_{A_1 A_2})+g[E_N(\varrho_{A_1A_2|B})]\leq E_{N,\mathrm{max}},
\end{equation}
where $E_{N,\mathrm{max}}$, the maximum value of $E_N$, is equal to $1/2$ for the two-qubit states, and the nondecreasing function $g$ is given by $g(x)=(3/2-\sqrt{1-2x^2}-\sqrt{1/4-x^2})/2$
\cite{Camalet}.  The monogamy inequality (\ref{EN}) has been calculated for our experimentally realized states in the Supplemental Material \cite{SM}.


Many familiar entanglement monotones satisfy monogamy inequalities of the form of Eq.~\eqref{EN} \cite{Camalet4}, but determining explicitly the corresponding function $g$ may not be always possible. Now we present another case for which this can be achieved.
For a qubit-qudit pure state $|\phi_{AB}\rangle$, the concurrence is defined by $C(|\phi_{AB}\rangle)=\sqrt{2(1-\tr \varrho_B^2)}$ \B{\cite{Wootters,CKW}}, where $\varrho_B=\tr_A (|\phi_{AB}\rangle\langle\phi_{AB}|)$ is the reduced density operator of system $B$.
It is generalized to mixed states via the convex roof extension \cite{Wootters,CKW}. For a three-qubit state $\varrho_{A_1A_2B}$, the following monogamy inequality holds
\begin{equation}\label{concurrence}
C(\varrho_{A_1 A_2})+\tilde{g}[C(\varrho_{A_1A_2|B})]\leq C_{\mathrm{max}},
\end{equation}
where $C_{\mathrm{max}}$, the maximum value of $C$, is equal to $1$ for the two-qubit states, and the nondecreasing function $\tilde g$ is given by $\tilde{g}(x)=(1-\sqrt{1-x^2})/2$
\cite{SM}.  The monogamy inequality (\ref{concurrence}) has also been calculated for our experimentally realized states in the Supplemental Material \cite{SM}.

\textit{Discussion and conclusion.---}
In this experiment, the visibility of the MZ interferometer is about 100:1 and the average fidelity \cite{Nielsen}  between the experimental states and theoretical states is $99.11\pm0.04\%$. Although the experimental states are not exactly pure states, the monogamy inequality (\ref{4}) still holds for experimental mixed states.
Moreover, the negativity in inequality \eqref{EN} has been evaluated for the mixed experimental states.
Besides the single-photon avalanche photon-diode is used to detect photons whose detection efficiency is~68\%. The detection events from the same pair are identified by a coincidence counter as long as they arrive within $\pm3$ns. In addition, the coincidence counts are about 1000$\text{s}^{-1}$ and we record clicks for 10s. There are many sources of the measurement uncertainty, such as counting statistics, detector efficiency, detector's dead time, timing uncertainty and alignment error of wave plates. However, the resulting uncertainty is dominated by counting statistics \cite{Hou Shun Poh2015}, which we have calculated via the Poissonian distribution and shown in the figures.

In summary, we have demonstrated the internal and external entanglement tradeoff in a photonic system with tunable entangled sources. This realization verifies the theoretical prediction that the entanglement 
between different degrees of freedom of a quantum single particle restricts its
entanglement with other particles. This property may have applications in quantum information, such as the construction of quantum communication network. On the other hand, this realization also can be generalized into other physical systems, such as NV centers, atoms, trapped ions, superconductor and so on. Particularly realization in a hybrid system, such as a photon and an atom, is more expected. For future research, one may experimentally demonstrate other monogamy inequalities, such as the inequality between local coherence and entanglement \cite{Fan,Camalet}, and the inequality between internal entanglement and external correlations \cite{Camalet4}.

This work is funded by the National Natural Science Foundation of China (Grants Nos.~11504253, 11575051, 11674306, 61590932 and 11734015), National Key R$\&$D Program (No. 2016YFA0301300 and No. 2016A0301700), Anhui Initiative in Quantum Information Technologies, the startup funding from Soochow University (Grant No. Q410800215) and the Hebei NSF (Grant No. A2016205215).


\begin{thebibliography}{42}%


\bibitem{mono1} R. Horodecki, P. Horodecki, M. Horodecki, and K. Horodecki, {\it Quantum entanglement}, Rev. Mod. Phys. \textbf{81}, 865 (2009).

\bibitem{mono2} C. H. Bennett, H. J. Bernstein, S. Popescu, and B. Schumacher, {\it Concentrating partial entanglement by local operations}, Phys. Rev. A \textbf{53}, 2046 (1996).

\bibitem{CKW} V. Coffman, J. Kundu, and W. K.Wootters, {\it Distributed entanglement}, Phys. Rev. A \textbf{61}, 052306 (2000).

\bibitem{Wootters} W. K. Wootters, {\it Entanglement of formation of an arbitrary state of two qubits}, Phys. Rev. Lett. \textbf{80}, 2245 (1998).

\bibitem{NCKW} T. J. Osborne and F. Verstraete, {\it General monogamy inequality for bipartite qubit entanglement}, Phys. Rev. Lett. \textbf{96}, 220503 (2006).

\bibitem{gaussian1} G. Adesso and F. Illuminati, {\it Continuous variable tangle, monogamy inequality, and entanglement sharing in Gaussian states of continuous variable systems}, New J. Phys. \textbf{8}, 15 (2006).

\bibitem{gaussian2} T. Hiroshima, G. Adesso, and F. Illuminati, {\it Monogamy inequality for distributed Gaussian entanglement}, Phys. Rev. Lett. \textbf{98}, 050503 (2007).

\bibitem{gaussian3} G. Adesso and F. Illuminati, {\it Strong Monogamy of Bipartite and Genuine Multipartite Entanglement: The Gaussian Case}, Phys. Rev. Lett. \textbf{99}, 150501 (2007).

\bibitem{gaussian4} G. Adesso and F. Illuminati, {\it Genuine multipartite entanglement of symmetric Gaussian states: Strong monogamy, unitary localization, scaling behavior, and molecular sharing structure}, Phys. Rev. A \textbf{78}, 042310 (2008).

\bibitem{SEntangle1} M. Christandl and A. Winter, {\it ``Squashed entanglement": An additive entanglement measure}, J. Math. Phys. (N.Y.) \textbf{45}, 829
(2004).

\bibitem{SEntangle2} M. Koashi and A. Winter, {\it Monogamy of quantum entanglement and other correlations}, Phys. Rev. A \textbf{69}, 022309
(2004).

\bibitem{negativity} Y.-C. Ou and H. Fan, {\it Monogamy inequality in terms of negativity for three-qubit states}, Phys. Rev. A \textbf{75}, 062308
(2007).

\bibitem{CREN} J. S. Kim, A. Das, and B. C. Sanders, {\it Entanglement monogamy of multipartite higher-dimensional quantum systems using convex-roof extended negativity}, Phys. Rev. A \textbf{79}, 012329 (2009).

\bibitem{Kim2} J. S. Kim, {\it Negativity and tight constraints of multiqubit entanglement}, Phys. Rev. A \textbf{97}, 012334 (2018).

\bibitem{Allen} G. W. Allen and D. A. Meyer, {\it Polynomial Monogamy Relations for Entanglement Negativity}, Phys. Rev. Lett. \textbf{118}, 080402 (2017).

\bibitem{SEOF} Y.-K. Bai, Y.-F. Xu, and Z. D. Wang, {\it General Monogamy Relation for the Entanglement of Formation in Multiqubit Systems}, Phys. Rev. Lett. \textbf{113}, 100503 (2014).

\bibitem{SEOF2} Y.-K. Bai, Y.-F. Xu, and Z. D. Wang, {\it Hierarchical monogamy relations for the squared entanglement of formation in multipartite systems}, Phys. Rev. A \textbf{90}, 062343 (2014).

\bibitem{SEOF3} T. R. de Oliveira, M. F. Cornelio, and F. F. Fanchini, {\it Monogamy of entanglement of formation}, Phys. Rev. A \textbf{89}, 034303 (2014).

\bibitem{bai2} Y.-K. Bai, M.-Y. Ye, and Z. D. Wang, {\it Entanglement monogamy and entanglement evolution in multipartite systems}, Phys. Rev. A \textbf{80}, 044301 (2009).



\bibitem{Kim3} J. S. Kim and B. C. Sanders, {\it Monogamy and polygamy for multi-qubit entanglement using R\'enyi entropy}, J. Phys. A: Math. Theor. \textbf{43}, 442305 (2010).

\bibitem{Renyi} M. F. Cornelio and M. C. de Oliveira, {\it Strong superadditivity and monogamy of the R\'enyi measure of entanglement}, Phys. Rev. A \textbf{81}, 032332 (2010).

\bibitem{Kim4} J. H. Choi and J. S. Kim, {\it Negativity and strong monogamy of multiparty quantum entanglement beyond qubits}, Phys. Rev. A \textbf{92}, 042307 (2015).

\bibitem{song} W. Song, Y.-K. Bai, M. Yang, M. Yang, and Z.-L. Cao, {\it General monogamy relation of multiqubit systems in terms of squared R\'enyi-$\alpha$ entanglement},  Phys. Rev. A \textbf{93}, 022306 (2016).

\bibitem{song2} G.-M. Yuan, W. Song, M. Yang, D.-C. Li, J.-L. Zhao, and Z.-L. Cao, {\it Monogamy relation of multi-qubit systems for squared Tsallis-q entanglement}, Sci. Rep. \textbf{6}, 28719 (2016).

\bibitem{exp} G. H. Aguilar, A. Vald\'{e}s-Hern\'{a}ndez, L. Davidovich, S. P. Walborn, and P. H. Souto Ribeiro, {\it Experimental Entanglement Redistribution under Decoherence Channels}, Phys. Rev. Lett. \textbf{113}, 240501 (2014).

\bibitem{exp2} O. Jim\'enez Far\'ias, A. Vald\'es-Hern\'andez, G. H. Aguilar, P. H. Souto Ribeiro, S. P. Walborn, L. Davidovich, X.-F. Qian and J. H. Eberly, {\it Experimental investigation of dynamical invariants in bipartite entanglement}, Phys. Rev. A \textbf{85}, 012314 (2012).


\bibitem{Gour} G. Gour, S. Bandyopadhyay, and B. C. Sanders, {\it Dual monogamy inequality for entanglement}, J. Math. Phys. \textbf{48}, 012108 (2007).

\bibitem{Yu} C.-s. Yu and H.-s. Song, {\it Entanglement monogamy of tripartite quantum states}, Phys. Rev. A \textbf{77}, 032329 (2008).

\bibitem{Sen} R. Prabhu, A. K. Pati, A. Sen(De), and U. Sen, {\it Relating monogamy of quantum correlations and multisite entanglement}, Phys. Rev. A \textbf{86}, 052337 (2012).

\bibitem{Sen2} R. Prabhu, A. K. Pati, A. Sen(De), and U. Sen, {\it Conditions for monogamy of quantum correlations: Greenberger-Horne-Zeilinger versus W  states}, Phys. Rev. A \textbf{85}, 040102(R) (2012).


\bibitem{Fei} X.-N. Zhu, and S.-M. Fei, {\it Entanglement monogamy relations of qubit systems}, Phys. Rev. A \textbf{90}, 024304 (2014).

\bibitem{Fei2} X.-N. Zhu, and S.-M. Fei, {\it Generalized monogamy relations of concurrence for N-qubit systems}, Phys. Rev. A \textbf{92}, 062345 (2015).

\bibitem{Fei3} X.-N. Zhu, X. Li-Jost, and S.-M. Fei, {\it Monogamy relations of concurrence for any dimensional quantum systems},
Quant. Inf. Process. \textbf{16}, 279 (2017).

\bibitem{ou2} Y.-C. Ou, H. Fan, and S.-M. Fei, {\it Proper monogamy inequality for arbitrary pure quantum states}, Phys. Rev. A \textbf{78}, 012311 (2008).



\bibitem{Regula} B. Regula, S. D. Martino, S. Lee, and G. Adesso, {\it Strong Monogamy Conjecture for Multiqubit Entanglement: The Four-Qubit Case}, Phys. Rev. Lett. \textbf{113}, 110501 (2014).

\bibitem{Regula2} B. Regula, A. Osterloh, and G. Adesso, {\it Strong monogamy inequalities for four qubits}, Phys. Rev. A \textbf{93}, 052338 (2016).

\bibitem{Osterloh} A. Osterloh, {\it Three-tangle of the nine classes of four-qubit states}, Phys. Rev. A \textbf{94}, 012323 (2016).

\bibitem{Luo} Y. Luo and Y. Li, {\it Monogamy of $\alpha$th power entanglement measurement in qubit systems}, Ann. Phys. (N. Y.) \textbf{362}, 511 (2015).

\bibitem{Luo2} Y. Luo, T. Tian, L.-H. Shao, and Y. Li, {\it General Monogamy of Tsallis-q Entropy Entanglement in Multiqubit Systems}, Phys. Rev. A \textbf{93}, 062340 (2016).

\bibitem{Winter} C. Lancien, S. Di Martino, M. Huber, M. Piani, G.
Adesso, and A. Winter, {\it Should Entanglement Measures be Monogamous or Faithful?} Phys. Rev. Lett. \textbf{117}, 060501 (2016).

\bibitem{Eltschka} C. Eltschka, A. Osterloh, and J. Siewert, {\it Possibility of generalized monogamy relations for multipartite
entanglement beyond three qubits}, Phys. Rev. A \textbf{80}, 032313 (2009).


\bibitem{Eltschka2} C. Eltschka and J. Siewert, {\it  Distribution of entanglement and correlations in all finite dimensions}, Quantum \textbf{2}, 64 (2018).

\bibitem{Eltschka3} C. Eltschka, F. Huber, O. G\"uhne, and J. Siewert, {\it Exponentially many entanglement and correlation constraints for multipartite
quantum states}, arXiv:1807.09165 (2018).

\bibitem{equa} C. Eltschka and J. Siewert, {\it Monogamy Equalities for Qubit Entanglement from Lorentz Invariance}, Phys. Rev. Lett. \textbf{114}, 140402 (2015).

\bibitem{equa2} G. Gour and Y. Guo, {\it Monogamy of entanglement without
inequalities}, Quantum \textbf{2}, 81 (2018).


\bibitem{Camalet} S. Camalet, {\it Monogamy Inequality for Any Local Quantum Resource and Entanglement}, Phys. Rev. Lett. \textbf{119}, 110503 (2017).

\bibitem{Camalet2} S. Camalet, {\it Simple state preparation for contextuality tests with few observables}, Phys. Rev. A \textbf{94}, 022106 (2016).

\bibitem{Camalet3} S. Camalet, {\it Monogamy inequality for entanglement and local contextuality}, Phys. Rev. A \textbf{95}, 062329 (2017).

\bibitem{Camalet4} S. Camalet, {\it Internal Entanglement and External Correlations of Any Form Limit Each Other}, Phys. Rev. Lett. \textbf{121}, 060504  (2018).


\bibitem{AS} M. Ku\'{s} and K. \.{Z}yczkowski, {\it Geometry of entangled states}, Phys. Rev. A \textbf{63}, 032307 (2001).

\bibitem{MEMS} F. Verstraete, K. Audenaert, and B. De Moor, {\it Maximally entangled mixed states of two qubits}, Phys. Rev. A \textbf{64}, 012316 (2001).


\bibitem{neg1}  K. \.{Z}yczkowski, P. Horodecki, A. Sanpera, and M. Lewenstein, {\it Volume of the set of separable states}, Phys. Rev. A \textbf{58}, 883 (1998).

\bibitem{neg2} G. Vidal and R. F. Werner, {\it Computable measure of entanglement}, Phys. Rev. A \textbf{65}, 032314 (2002).

\bibitem{ER} V. Vedral and M. Plenio, {\it Entanglement measures and purification procedures}, Phys. Rev. A \textbf{57}, 1619 (1998).

\bibitem{SM} See Supplemental Material for detailed explanations of other monogamy inequalities.


\bibitem{P.G.Kwiat1999}
P. G. Kwiat, E. Waks, A. G. White, I. Appelbaum, and P. H. Eberhard, {\it Ultrabright source of polarization-entangled photons}, Phys. Rev. A \textbf{60}, R773 (1999).

\bibitem{Daniel F.V.James2001}
D. F. V. James, P. G. Kwiat, W. J. Munro and A. G. White, {\it Measurement of qubits}, Phys. Rev. A \textbf{64}, 052312 (2001).


\bibitem{Nielsen}
M. A. Nielsen and I. L. Chuang, {\it Quantum computation and quantum information} (Cambridge University Press, Cambridge, 2000).

\bibitem{Hou Shun Poh2015}
H. S. Poh, S. K. Joshi, A. Cer\.{e}, A. Cabello and C. Kurtsiefer, {\it Approaching Tsirelson's bound in a photon pair experiment}, Phys. Rev. Lett. \textbf{115}, 180408 (2015).	

\bibitem{Fan} Z. Xi, Y. Li and H. Fan, {\it Quantum coherence and correlations in quantum system},  Sci. Rep. \textbf{5}, 10922 (2015).


\end{thebibliography}

\begin{thebibliography}{42}%
\bibitem{SMEMS} F. Verstraete, K. Audenaert, and B. De Moor, {\it Maximally entangled mixed states of two qubits}, Phys. Rev. A \textbf{64}, 012316 (2001).


\bibitem{SCamalet} S. Camalet, {\it Monogamy Inequality for Any Local Quantum Resource and Entanglement}, Phys. Rev. Lett. \textbf{119}, 110503 (2017).

\bibitem{SWootters} W. K. Wootters, {\it Entanglement of formation of an arbitrary state of two qubits}, Phys. Rev. Lett. \textbf{80}, 2245 (1998).
\end{thebibliography}

%

%

\newpage
\onecolumngrid
\setcounter{page}{1}
\renewcommand{\thepage}{Supplemental Material --\arabic{page}/5}
\setcounter{equation}{0}
\setcounter{figure}{0}
\renewcommand{\theequation}{S\arabic{equation}}
\renewcommand{\thefigure}{S\arabic{figure}}

\section{SUPPLEMENTAL MATERIAL}

\subsection{I. Inequality (3) for the negativity $E_N$}
If we use the negativity $E_N$ to quantify the internal entanglement between $A_1$ and $A_2$, the inequality (3) in the main text becomes to
\begin{eqnarray}
E_N(\varrho_{A_1 A_2})+E_N'(|\psi\rangle_{A_1 A_2 |B})\leq\frac{1}{2},\label{s4}
\end{eqnarray}
where $E_N(\varrho_{A_1 A_2})=(\|\varrho_{A_1A_2}^{T_{A_2}}\|-1)/2$, and $E_{N,\mathrm{max}}$, the maximum value of $E_N(\varrho_{A_1 A_2})$, is equal to $1/2$ for two-qubit state $\varrho_{A_1 A_2}$. The external entanglement $E_N'(|\psi\rangle_{A_1 A_2 |B})$ is
\begin{equation}\label{}
    E_N'(|\psi\rangle_{A_1 A_2 |B})=\frac{1}{2}-E_N(\varrho_{A_1 A_2}'),\label{s5}
\end{equation}
with
\begin{eqnarray}
E_N(\varrho_{A_1 A_2}')&=&\max_U E_N(U\varrho_{A_1 A_2}U^\dag)\nonumber\\
&=&  \max\{0,\frac{1}{2}\sqrt{(\lambda_1-\lambda_3)^2+(\lambda_2-\lambda_4)^2}-\frac{\lambda_2}{2}-\frac{\lambda_4}{2}\}\label{s6}
\end{eqnarray}
where $U$ denotes the unitary operators of $A$, $\varrho_{A_1 A_2}'$ the density
operator corresponding to the maximum over $U$,
and $\{\lambda_i\}$ are the eigenvalues of $\varrho_{A_1 A_2}$ in nonascending order
\cite{SMEMS}.

\begin{figure}[htbp]
\centering
\begin{minipage}[t]{0.48\textwidth}
\centering
\includegraphics[scale=0.6]{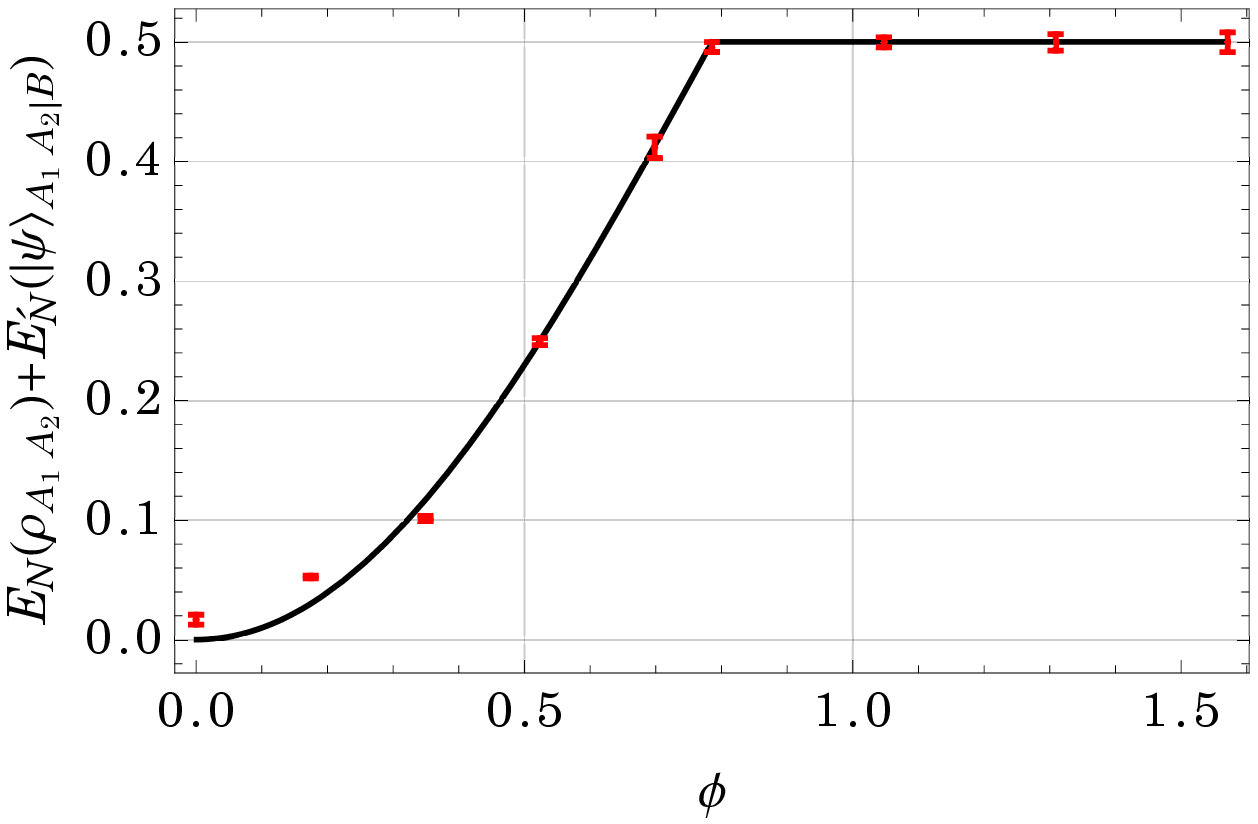}
\end{minipage}
\begin{minipage}[t]{0.48\textwidth}
\centering
\includegraphics[scale=0.6]{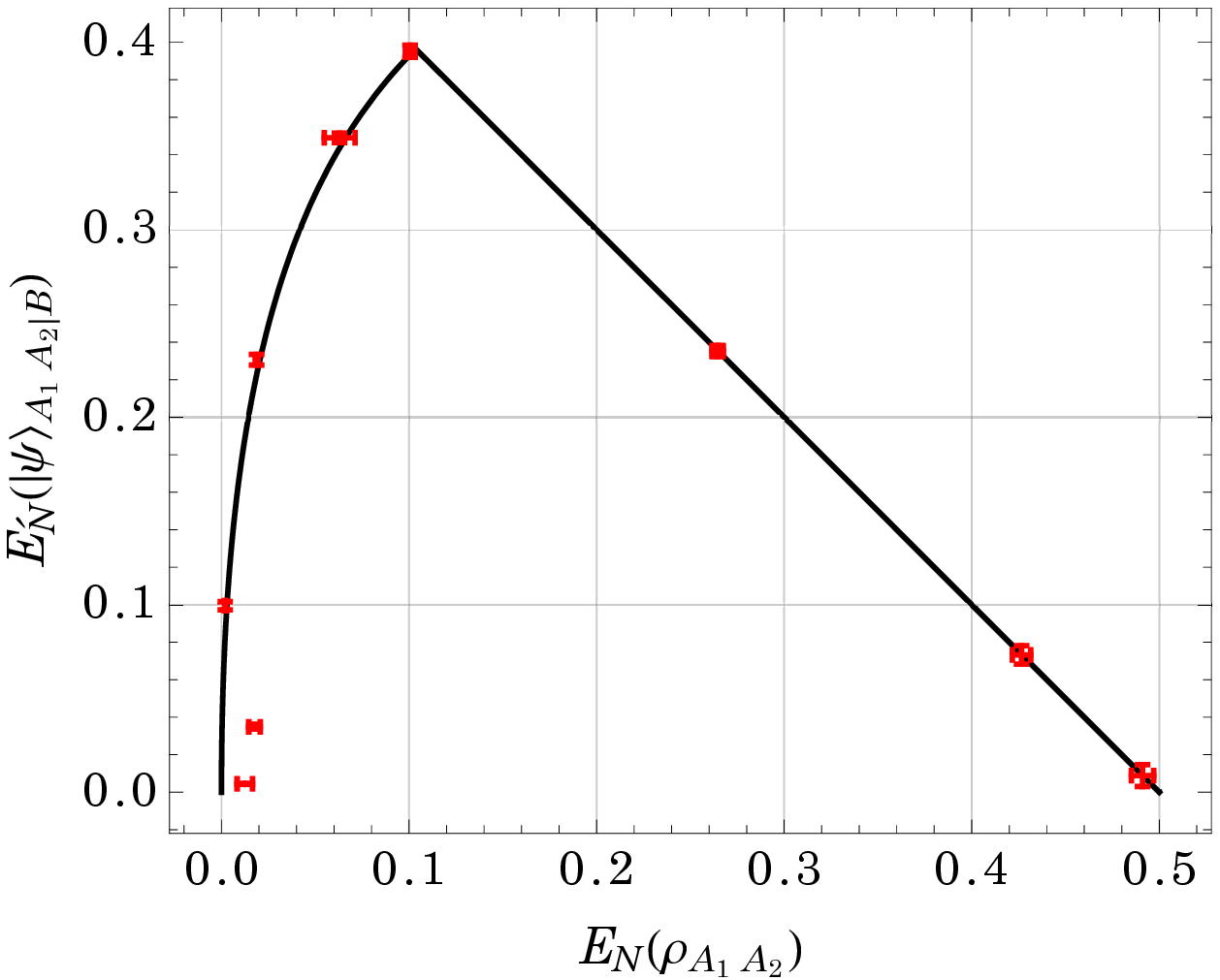}
\end{minipage}
\caption{We use the negativity $E_N(\varrho_{A_1 A_2})$ and $E_N'(|\psi\rangle_{A_1 A_2 |B})$ to quantify the entanglement among these three qubits. The red dots are experimental results and the lines are theoretical predictions.}\label{figS1}
\end{figure}

Now we consider a class of three-qubit pure states with one parameter $\phi$,
\begin{eqnarray}
|\psi\rangle=\cos\phi|110\rangle+\sin\phi\frac{|01\rangle+|10\rangle}{\sqrt{2}}|1\rangle.\label{}
\end{eqnarray}
Based on Eqs. (\ref{s4})-(\ref{s6}), one can obtain its internal and external entanglement measured by the negativity,
\begin{eqnarray}
E_N(\varrho_{A_1 A_2})&=& \frac{1}{4} \sqrt{3+\cos(4\phi)}  -\frac{1}{2} \cos^2\phi,\\
E_N'(|\psi\rangle_{A_1 A_2 |B})&=& \frac{1}{2}+\frac{1}{2} \min\{\cos^2\phi,\sin^2\phi\}-\frac{1}{4} \sqrt{3+\cos(4\phi)}  .
\end{eqnarray}
The theoretical and experimental results have been shown in Fig. \ref{figS1}. We can see that all the sums of internal and external entanglement are bounded by $1/2$, i.e., the inequality (\ref{s4}) always holds.

\subsection{II. The monogamy inequality (9)  involving only $E_N$}
In Ref. \cite{SCamalet}, the author also presented a monogamy inequality involving only one entanglement monotone, the negativity $E_N$. For a bipartite state $\varrho_{AB}$, the negativity is defined by $E_N(\varrho_{AB})=(\|\varrho_{AB}^{T_B}\|-1)/2$, where $\|\cdot\|$ is the trace norm and $T_B$ is the partial transpose 
with respect to system $B$.
For a three qubit pure state $\ket{\psi}_{A_1A_2B}$, the monogamy inequality is
\begin{equation}\label{sEN}
E_N(\varrho_{A_1 A_2})+g[E_N(\ket{\psi}_{A_1A_2|B})]\leq \frac{1}{2},
\end{equation}
where $E_{N,\mathrm{max}}$, the maximum value of $E_N(\varrho_{A_1 A_2})$, is equal to $1/2$ for two-qubit state $\varrho_{A_1 A_2}$, and the nondecreasing function
$g$ is given by
\begin{equation}\label{gfunction}
    g(x)=\frac{3}{4}-\frac{\sqrt{1-2x^2}}{2}-\frac{\sqrt{1-4x^2}}{4},
\end{equation}
when the number of nonzero eigenvalues of $\varrho_{A_1 A_2}$ is equal to 2 \cite{SCamalet}.

\begin{figure}[htbp]
\centering
\begin{minipage}[t]{0.48\textwidth}
\centering
\includegraphics[scale=0.6]{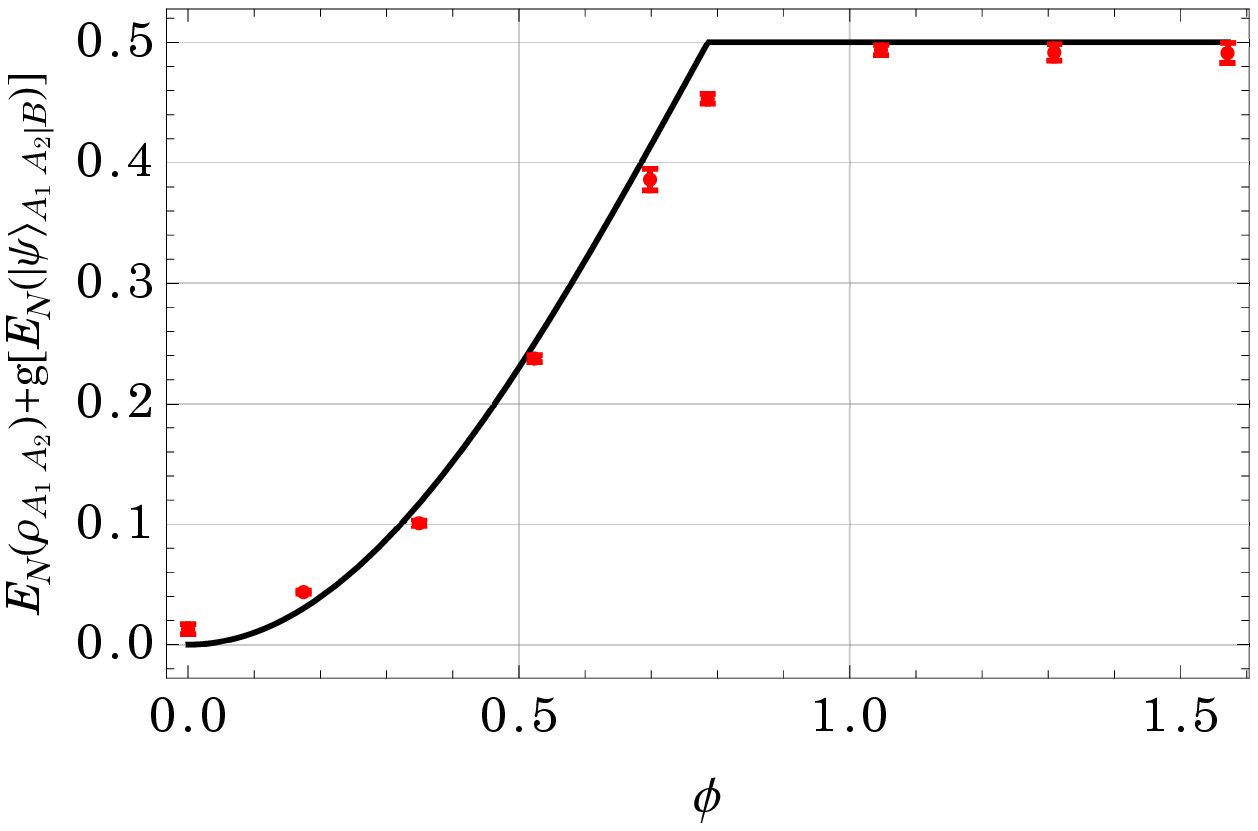}
\end{minipage}
\begin{minipage}[t]{0.48\textwidth}
\centering
\includegraphics[scale=0.6]{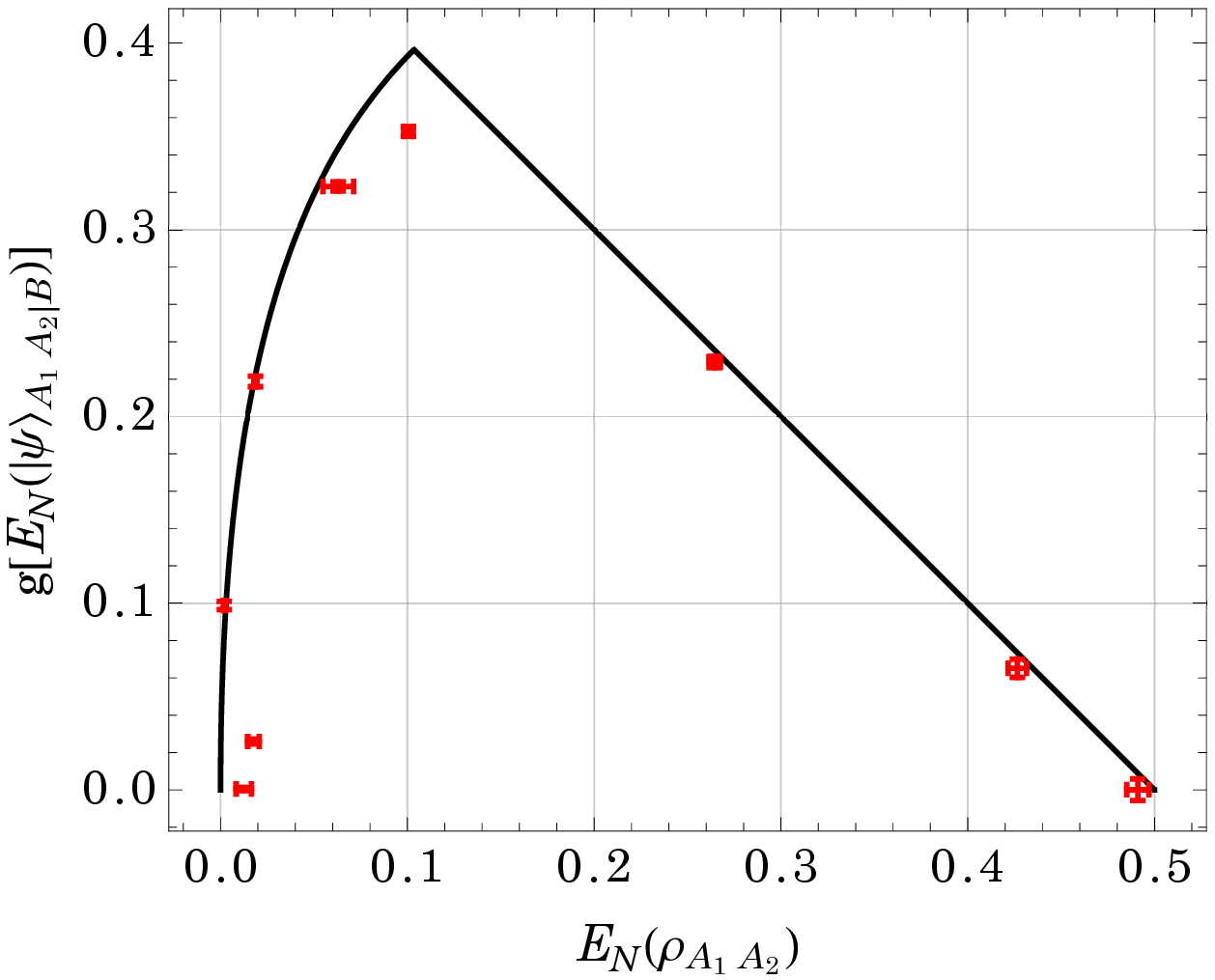}
\end{minipage}
\caption{In order to use the same measure, $E_N(\varrho_{A_1 A_2})$ and $E_N(\ket{\psi}_{A_1A_2|B})$, to quantify the entanglement, we employ the function $g$. Note that $g$ is 
not shown in the Figure. The red dots are experimental results and the lines are theoretical predictions.}\label{figS2}
\end{figure}

Now we consider a class of three-qubit pure states with one parameter $\phi$,
\begin{eqnarray}
|\psi\rangle=\cos\phi|110\rangle+\sin\phi\frac{|01\rangle+|10\rangle}{\sqrt{2}}|1\rangle.\label{}
\end{eqnarray}
Based on Eqs. (\ref{sEN}) and (\ref{gfunction}), one can obtain its internal and external entanglement measured by the negativity,
\begin{eqnarray}
E_N(\varrho_{A_1 A_2})&=&  \frac{1}{4} \sqrt{3+\cos(4\phi)}  -\frac{1}{2} \cos^2\phi  ,\\
g[E_N(\ket{\psi}_{A_1A_2|B})]&=& \frac{3}{4}-\frac{\sqrt{1-2\cos^2\phi\sin^2\phi}}{2}-\frac{\sqrt{1-4\cos^2\phi\sin^2\phi}}{4}.
\end{eqnarray}
The theoretical and experimental results have been shown in Fig. \ref{figS2}. We can see that all the  sums of internal and external entanglement are bounded by $1/2$, i.e., the inequality (\ref{sEN}) always holds.

\subsection{III. The monogamy inequality (10) involving only $C$}
\textit{Proposition 1.} For a three-qubit state $\varrho_{A_1A_2B}$,
the internal entanglement $C(\varrho_{A_1A_2})$ and the external entanglement
$C(\varrho_{A_1A_2|B})$,
as quantified by the concurrence, obey the monogamy relation
\begin{equation}\label{s12}
C(\varrho_{A_1A_2})+\tilde{g}[C(\varrho_{A_1A_2|B})]\leq 1,
\end{equation}
where $1$ is the maximal value of $C$ for two-qubit states  and
\begin{equation}\label{s13}
   \tilde{g}(x)=\frac{1-\sqrt{1-x^2}}{2}.
\end{equation}

\textit{Proof.}~For a pure three-qubit state, the external entanglement is given by
\begin{equation}\label{}
 C(\ket{\psi}_{A_1A_2|B})=\sqrt{2(1-\tr\varrho_{A_1A_2}^2)}=2\sqrt{\lambda_1(1-\lambda_1)},
\end{equation}
with $\lambda_1$ being the maximal eigenvalue of $\varrho_{A_1A_2}$.
Since the above function of $\varrho_{A_1A_2}$ is concave and the concurrence is defined via the convex roof extension for mixed states, the external entanglement obeys
\begin{equation}\label{}
 C(\varrho_{A_1A_2|B}) \le 2\sqrt{\lambda_1(1-\lambda_1)},
\end{equation}
in the general case.
The internal entanglement can be obtained by the formula $C(\varrho_{A_1A_2})=\max\{0,\sigma_1-\sigma_2-\sigma_3-\sigma_4\}$ with $\{\sigma_i\}$ being the square roots of eigenvalues of $\varrho_{A_1A_2}\sigma_y\otimes\sigma_y \varrho_{A_1A_2}^*\sigma_y\otimes\sigma_y$ in decreasing order  \cite{SWootters}. In the bipartite partition $A_1A_2|B$, the internal
entanglement can be maximized via two-qubit unitary transformations on $A_1A_2$, and the following relation holds
\begin{eqnarray}
    C(\varrho_{A_1A_2})&\leq& \max_U C(U\varrho_{A_1A_2}U^\dagger),
\end{eqnarray}
where the equality is satisfied for the so-called maximally entangled mixed state (MEMS) $\varrho'_{A_1A_2}$ \cite{SMEMS}. In the
case of two-qubit MEMSs \cite{SMEMS}, its concurrence is
$C(\varrho'_{A_1A_2})=\max\{0, \lambda_1-\lambda_3
-2\sqrt{\lambda_2\lambda_4} \} \le \lambda_1$, and hence
$C(\varrho_{A_1A_2}) \le \lambda_1$.
Therefore, we have
\begin{eqnarray}
 C(\varrho_{A_1A_2})+\tilde{g}[C(\varrho_{A_1A_2|B})]&\leq& \lambda_1+\tilde{g}\Big(2\sqrt{\lambda_1(1-\lambda_1)}\Big)\nonumber\\
 &=&\lambda_1+\frac{1-\sqrt{1-4\lambda_1(1-\lambda_1)}}{2}\nonumber\\
 &=&\lambda_1+\frac{1-(2\lambda_1-1)}{2}\nonumber\\
 &=&1,
\end{eqnarray}
where $1/2\leq\lambda_1\leq1$ has been used. Thus, we obtain the monogamy relation (\ref{s12}).  \hfill $\square$

\begin{figure}[htbp]
\centering
\begin{minipage}[t]{0.48\textwidth}
\centering
\includegraphics[scale=0.6]{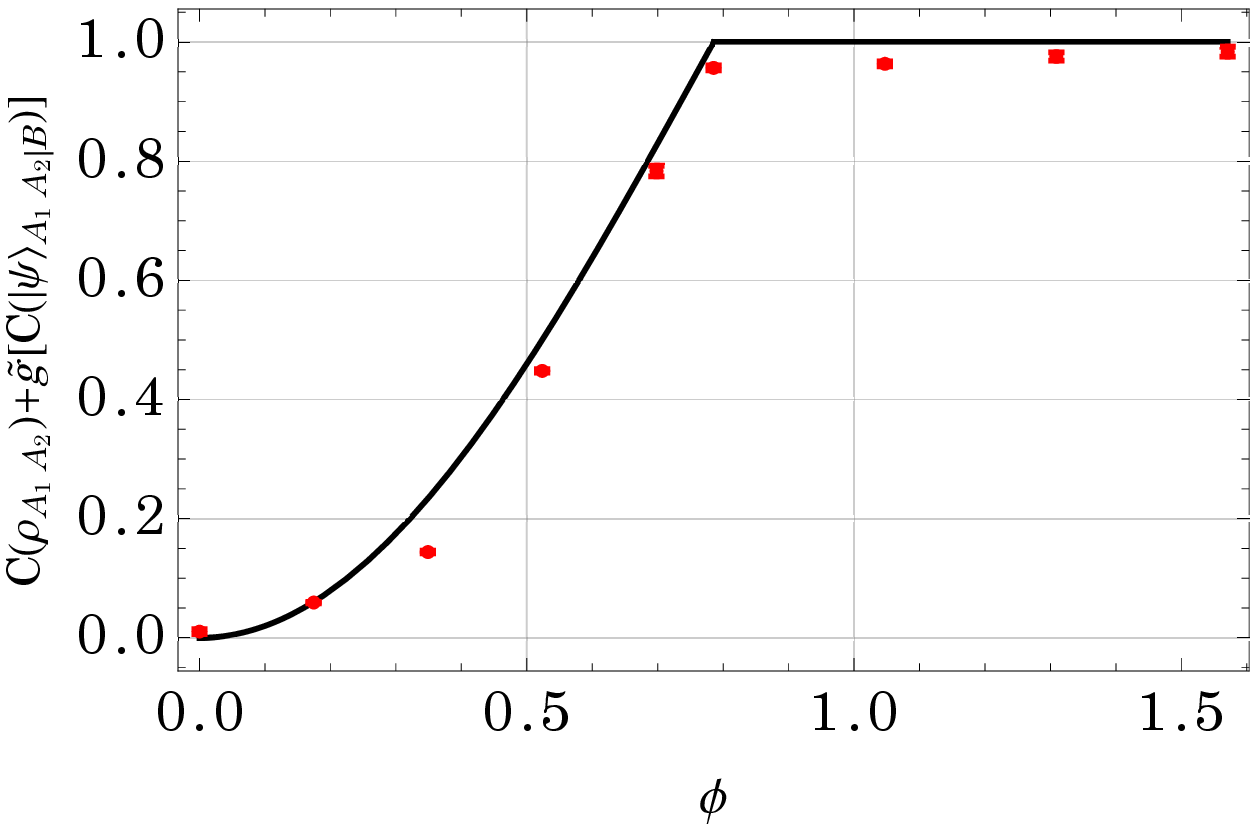}
\end{minipage}
\begin{minipage}[t]{0.48\textwidth}
\centering
\includegraphics[scale=0.6]{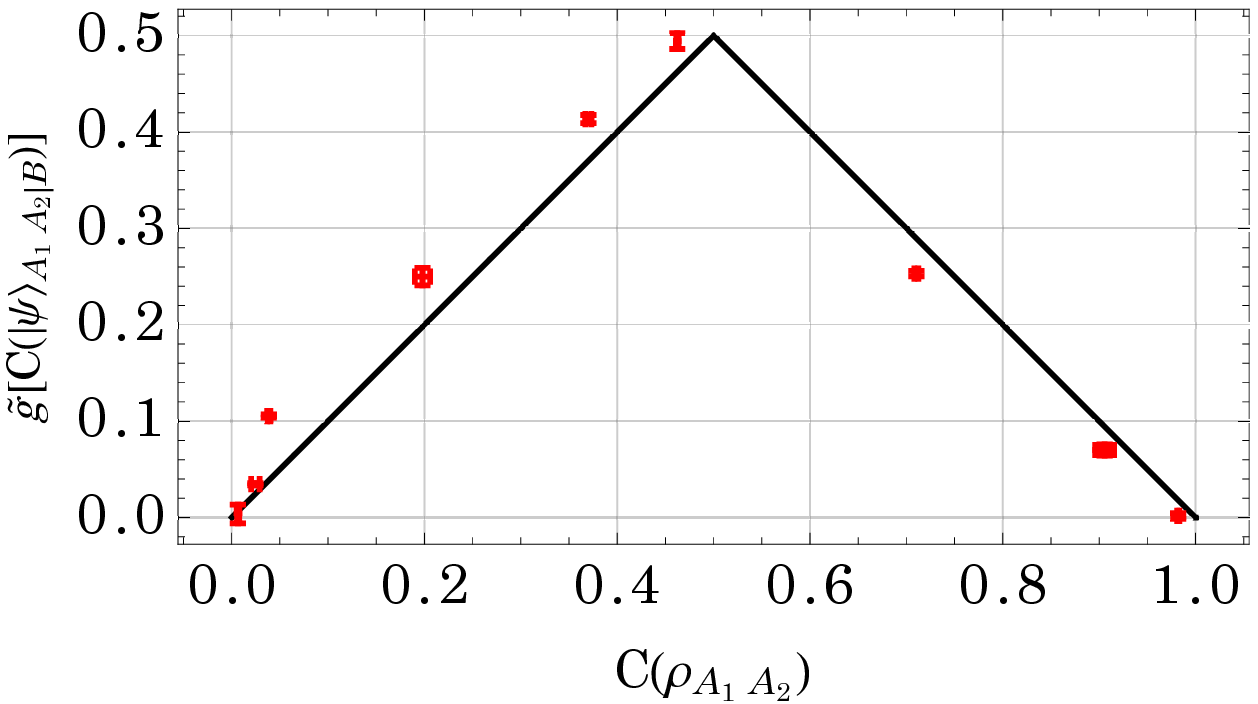}
\end{minipage}
\caption{We use the concurrence $C$ to quantify the internal and external entanglement simultaneously so we introduce the function $\tilde{g}$ which is not shown in the Figure. The red dots are experimental results and the lines are theoretical predictions.}\label{figS3}
\end{figure}

Now we consider a class of three-qubit pure states with one parameter $\phi$,
\begin{eqnarray}
|\psi\rangle=\cos\phi|110\rangle+\sin\phi\frac{|01\rangle+|10\rangle}{\sqrt{2}}|1\rangle.\label{}
\end{eqnarray}
One can obtain its internal and external entanglement measured by
concurrence,
\begin{eqnarray}
C(\varrho_{A_1 A_2})&=&   \sin^2\phi,\\
\tilde{g}[C(\ket{\psi}_{A_1A_2|B})]&=& 1-\lambda_1,
\end{eqnarray}
with
\begin{equation}\label{}
    \lambda_1=\max\{\cos^2\phi,\sin^2\phi\}.
\end{equation}
One can see that if $\lambda_1=\sin^2\phi$ (i.e., $\sin^2\phi\geq\cos^2\phi$), then $C(\varrho_{A_1 A_2})+\tilde{g}[C(\ket{\psi}_{A_1A_2|B})]=1$ holds.
The theoretical and experimental results have been shown in Fig. \ref{figS3}. We can see that all the  sums of internal and external entanglement are bounded by 1, i.e., the inequality (\ref{s12}) always holds.

\subsection{IV. Quantum state tomography}

We performed tomography to nine states we prepared. Their density matrices can be described in Eq. (8) of the main text. The $\theta$ is fixed at $45^{\circ}$ but the $\phi$ is changing. Following are their density matrices that are obtained by maximum likelihood estimation. We only labeled the values of $\phi$ to distinguish different states and also listed the fidelity of them.

\begin{table}[h]
\renewcommand\arraystretch{1.3}
\centering
\caption{\label{Table:1} Fidelities for the quantum state Eq. (8) of the main text. The $\theta$ is fixed at $45^{\circ}$ but the $\phi$ is changing from $0^{\circ}$ to $90^{\circ}$. The average fidelity of these states is $99.11\pm0.04\%$.}

\begin{tabular}{c | c c c c c}
\hline
\hline
$\phi$ & $0^{\circ}$ &  $15^{\circ}$ & $30^{\circ}$ & $45^{\circ}$ & $50^{\circ}$    \\
\hline
Fidelity &  $99.05\pm0.11\%$ &  $99.16\pm0.07\%$ & $99.51\pm0.02\%$ & $99.16\pm0.05\%$ & $98.94\pm0.05\%$   \\
\hline
\hline
$\phi$ & $60^{\circ}$ & $70^{\circ}$ & $80^{\circ}$ & $90^{\circ}$ & average   \\
\hline
Fidelity  & $99.10\pm0.06\%$ & $98.77\pm0.03\%$ & $98.20\pm0.12\%$ & $99.39\pm0.05\%$ & $99.11\pm0.04\%$  \\
\hline

\end{tabular}

\end{table}

Although the experimental states are not exactly pure states, the monogamy inequality (3) in the main text still holds for experimental mixed states. Suppose that the experimentally realized tripartite state is $\varrho_{A_1 A_2 B}$, thus the external entanglement is defined by the convex roof,
\begin{eqnarray}
E_F'(\varrho_{A_1 A_2|B})=\inf_{\{p_i,|\psi_i\rangle_{A_1 A_2 B}\}}\sum_i p_i E_F'(|\psi_i\rangle_{A_1 A_2 |B}).
\end{eqnarray}
We assume that  $\varrho_{A_1 A_2 B}=\sum_j p_j|\psi_j\rangle\langle\psi_j|$ is the optimal decomposition for $\varrho_{A_1 A_2 B}$ to achieve the above infimum. Therefore,
\begin{eqnarray}
E_F(\varrho_{A_1 A_2})+E_F'(\varrho_{A_1 A_2|B})&=&E_F(\varrho_{A_1 A_2})+\sum_j p_j E_F'(|\psi_j\rangle_{A_1 A_2 |B})\nonumber\\
&=&E_F(\varrho_{A_1 A_2})+1-\sum_j p_j E_F(U_j \varrho_{j,A_1A_2}U_j^\dag)\nonumber\\
&\leq&E_F(\varrho_{A_1 A_2})+1-\sum_j p_j E_F(V \varrho_{j,A_1A_2}V^\dag)\nonumber\\
&\leq&E_F(\varrho_{A_1 A_2})+1-E_F(V\sum_j p_j\varrho_{j,A_1A_2}V^\dag)\nonumber\\
&=&E_F(\varrho_{A_1 A_2})+1-E_F(\varrho_{A_1 A_2}')\nonumber\\
&\leq&1,
\end{eqnarray}
where $ \varrho_{j,A_1A_2}=\tr_B |\psi_j\rangle\langle\psi_j| $, $E_F'(|\psi_j\rangle_{A_1 A_2 |B})=1-E_F(\varrho_{j,A_1 A_2}')$,  $E_F(\varrho_{j,A_1 A_2}')=\max_U E_F(U\varrho_{j,A_1 A_2}U^\dag)=E_F(U_j\varrho_{j,A_1 A_2}U_j^\dag)$, $E_F(\varrho_{A_1 A_2}')=\max_U E_F(U\varrho_{A_1 A_2}U^\dag)=E_F( V\varrho_{A_1 A_2}V^\dag)$, the first inequality holds since $E_F(U_j \varrho_{j,A_1A_2}U_j^\dag)=\max_U E_F(U\varrho_{j,A_1 A_2}U^\dag)\geq E_F(V \varrho_{j,A_1A_2}V^\dag)$, and the second inequality is from the convex property of $E_F$. This result also follows from the fact that $E'_F$ is a concave function of $\varrho_{A1A2}$, as shown by the Proposition 2 of the supplemental material of
Ref. \cite{SCamalet}.

\end{document}